\documentclass[preprint,aps]{revtex4}

\usepackage{graphicx,subfigure}

\newcommand{\be}{\begin{equation}}
\newcommand{\ee}{\end{equation}}
\newcommand{\bea}{\begin{eqnarray}}
\newcommand{\eea}{\end{eqnarray}}
\newcommand{\bd}{\begin{displaymath}}
\newcommand{\ed}{\end{displaymath}}

\newcommand{\op}{ \oplus_q  }
\newcommand{\om}{ \ominus_q  }

\newcommand{\qi }{ q^{-1} }

\newcommand{\dd}{D^{q^{2}}_u}

\newcommand{\lb }{ \left( }
\newcommand{\rb }{ \right ) }

\begin{document}


\title{
New $q-$Hermite polynomials: characterization, operator algebra and associated coherent states
}

\author{ Won Sang Chung }
\email{mimip4444@hanmail.net}

\author{ Mahouton Norbert Hounkonnou${}^b$}
\email{norbert.hounkonnou@cipma.uac.bj}

\author{ Arjika Sama${}^b$ }
\email{rjksama2008@gmail.com}

\affiliation{
Department of Physics and Research Institute of Natural Science, College of Natural Science, Gyeongsang National University, Jinju 660-701, Korea
}
\vspace{1cm}

\affiliation{
${}^b$International Chair of Mathematical Physics
and Applications
(ICMPA-UNESCO Chair), University of
Abomey-Calavi,
072 B. P.: 50 Cotonou, Republic of Benin}

\date{\today}

\begin{abstract}
This paper addresses a construction of new $q-$Hermite polynomials with a full characterization of their main properties and corresponding raising and lowering operator algebra. The three-term recursive relation as well as the second-order differential equation obeyed by these new polynomials are explicitly derived. Relevant operator actions, including the eigenvalue problem of the deformed oscillator and the self-adjointness of the related position and momentum operators, are investigated and analyzed. The associated coherent states  are constructed and discussed with an explicit resolution of the induced moment problem.



\end{abstract}

\maketitle

\section{Introduction}

In the last decade, quantum algebras and quantum groups have been the subject of intensive research in several physics and mathematics fields. Quantum groups or q-deformed Lie algebra implies some specific deformation of classical Lie algebra. From the mathematical point of view, it is a non-commutative associative Hopf algebra. The structure and representation theory of quantum groups have been developed extensively by Jimbo \cite{Jimbo} and Drinfeld \cite{Drinfeld}. 

The q-deformation of the oscillator algebra was firstly accomplished by Arik and Coon \cite {AC} and lately accomplished by Macfarlane \cite{Macfarlane} and Biedenharn \cite{Biedenharn} by using the q-calculus which was originally introduced by Jackson in the early 20th century \cite{Jackson1}. In the study of the basic hypergeometric function Jackson invented the Jackson q- derivative and q-integral. Jackson's pioneering research enabled theoretical physicists and mathematician to study the new physics or mathematics related to the q-calculus. Much was accomplished in this direction and work is under way to find the meaning of the q-deformed theory.

It is well known that the Hermite polynomial is related to the realization of an ordinary harmonic oscillator algebra. Recently Galetti \cite{Galetti} has shown that a similar procedure can be carried out in the case of the three term recurrence relation for Rogers Szeg\"o polynomials and the Jackson q- derivative. This technique furnished new realizations of the q-deformed algebra associated with the q-deformed harmonic oscillator, which obey, well known and spread in the literature, commutation relations. 

The present investigation aims at giving a new q-Hermite polynomials with a full characterization of their main properties and corresponding raising and lowering operator algebra. We will derive the three-term recurrence relation as well as the second-order differential equation obeyed by these new polynomials.

 The paper is organized as follows. As a matter of clarity, we present in Section II a  brief review of known results on Hermite polynomials and $q-$Rogers Szeg\"o polynomials. In Section III, we present the new $q-$Hermite polynomials and discuss its associated deformed oscillator algebra. In Section IV, we discuss the relevant operator properties such as the eigenvalue problem and the self-adjointness of the related position and momentum operators. In Section V, we construct the associated coherent states.

\section{Quick overview on  the Hermite polynomials and $q-$Rogers Szeg\"o polynomials}
For the clarity of our development, let us briefly sketch and discuss in this section main relevant results on Hermite polynomials and $q-$Rogers Szeg\"o polynomials.
\subsection{On the Hermite polynomials and Rogers Szeg\"o polynomials}
The classical Hermite polynomials are defined
 as orthogonal polynomials  satisfying the
three-term recursion relation
\be\label{4Hermttr}
H_{n+1}(z)=2z H_n(z)-2n H_{n-1}(z)
\ee
and  the first order differential equation
\be \label{4Hermdiff}
\frac{d}{dz}H_n(z)= 2n H_{n-1}(z).
\ee
Combining the equation
 (\ref{4Hermdiff}) and  (\ref{4Hermttr}), we obtain
\be
\label{rai}
H_{n+1}(z)= \Big(2z-\frac{d}{dz}\Big)H_n(z)
\ee
and naturally from (\ref{4Hermdiff}) and (\ref{rai})  introduce  the
lowering (annihilation) and raising (creation) operators (see
\cite{Galetti,Dezire}) as
\bea
\hat{a}_-&=& \frac{1}{2}\frac{d}{dz},\\
\hat{a}_+&= &2z-\frac{d}{dz}.
\eea
The set of Hermite
 polynomials can be then generated by the
application of the creation operator to the first polynomial
$H_0(z)=1$, i.e.,
\be
H_n(z)= \hat{a}_+^n H_0(z).
\ee
These polynomials are solutions of the
following second order
differential equation:
\be
\label{second}
\Big(\frac{d^2}{dz^2}-
2z\frac{d}{dz}+
2n\Big)H_n(z)=0.
\ee
Furthermore, defining a
number operator as
\be
\hat{n}:= \hat{a}_+\hat{a}_-,
\ee
one can readily check that  the following canonical commutation relation
\be \label{4HarmAlg}
[ \hat{a}_-,\;\hat{a}_+]={\bf 1},
\ee
as well as the peculiar expressions
\be
 [ \hat{n},\;\hat{a}_-]=-\hat{a}_- ,\qquad [ \hat{n},\;\hat{a}_+]=\hat{a}_+,
\ee
are verified
although the operators $\hat{a}_-$  and $\hat{a}_+$ are not in the same form as the
usual creation and annihilation
operators associated
with the quantum mechanical harmonic
oscillator. Here, the construction of the
 raising and lowering
operators stem from
the two basic relations (\ref{4Hermttr}) and
 (\ref{4Hermdiff}) satisfied by the Hermite polynomials,
i. e. the three-term recursion relation
and the differentiation
relation, respectively.

Considering the usual Fock space
$\mathcal{F}:=\big\{|n\rangle,\;
\langle m|n\rangle =n!\delta_{mn},\,\;n=0,1,2,\cdots\big\}$,  the
 raising operator $\hat{a}_+$ and the
lowering operator $\hat{a}_-$ satisfy the following relations
\bea
&&|n\rangle = \hat{a}_+^n|0\rangle,\cr
&&\hat{a}_-|0\rangle =0
\eea
and their actions  on $|n\rangle$ are given by
\bea
&&\hat{a}_+|n\rangle = |n+1\rangle,\cr
&&\hat{a}_-|n\rangle = |n-1\rangle.
\eea
To conclude this subsection, let us emphasize an alternative construction of raising and lowering operators proposed in
 \cite{Jagannathan&Sridhar10}, which
 considers the  sequence of Rogers Szeg\"o polynomials
\be
\psi_n(z)=\frac{1}{\sqrt{n!}}{\bf h}_n(z),
\ee
where
\be
{\bf h}_n(z):=(1+z)^n =\sum_{k=0}^{n}\left(\begin{array}{c}n\\k\end{array}\right) z^k,
\ee
obeying the  relations
\bea
\frac{d}{dz}\psi_n(z)&=&\sqrt{n}\,\psi_{n-1}(z),\\
(1+z)\psi_n(z) &=&\sqrt{n+1}\, \psi_{n+1}(z),\label{4Hermttr1}\\
(1+z)\frac{d}{dz}\psi_n(z)&=&n\psi_{n}(z),\label{4Hermdiff1}\\
\frac{d}{dz}\left((1+z)\psi_n(z)\right)&=&(n+1)\psi_n(z).
\eea
It comes, by analogy to the work done by Galleti \cite{Galetti}, that  the  raising, lowering and number
operators can be deduced in the following forms \cite{Jagannathan&Sridhar10}:
\be
\hat{a}_+=(1+z),\qquad \hat{a}_-=\frac{d}{dz}, \qquad \hat{n}= (1+z)\frac{d}{dz},
\ee
 respectively, and the set
$\{\psi_n(z)\;|n=0, 1, 2, \cdots\}$ forms a basis for the
Bargman-Fock realization of the harmonic oscillator.
%
%
\subsection{On the $q-$Rogers-Szeg\"o polynomials }\label{4qHosc}
Mimicking the procedure proposed by Jagannathan and
Sridhar \cite{Jagannathan&Sridhar10}, one can  construct the creation, annihilation and
number operators  from the three-term recurrence relation and the
$q-$difference equation founding the Rogers-Szeg\"o
polynomials. Recall that
 the $q-$deformed Rogers-Szeg\"o polynomials   are defined as \cite{Dezire}
\be
\label{deformed}
H_n(z;q):=\sum_{k=0}^{n}\left[\begin{array}{c}n\\k\end{array}\right]_q z^k,\quad n=0,1,2\cdots
\ee
 satisfying a three-term recursion relation
\be \label{4qRogersttr}
H_{n+1}(z;q)=(1+z)H_n(z;q)-z(1-q^n)H_{n-1}(z;q).
\ee
The action of the $q-$Jackson's derivative   on the polynomials
(\ref{deformed}) is given by
\be
D_z^q H_n(z;q)= [n]_q H_{n-1}(z;q),
\ee
where $D_z^q$ is defined are follows
\bea
\label{jackson}
D_z^q f(x)=\frac{f(z)-f(qz)}{(1-q)z}.
\eea
In the limit case $q\to 1$, the Rogers-Szeg\"o polynomial (\ref{deformed}) of degree $n,$ ($n= 0, 1, 2, \cdots$), well converges to
\be
{\bf h}_n(z) =\sum_{k=0}^{n}\left(\begin{array}{c}n\\k\end{array}\right) z^k
\ee
as expected.
By defining the polynomials
\bea\label{4qRogersnorm}
\psi_n(z;q)=\frac{1}{\sqrt{[n]_q!}}H_n(z)
=\frac{1}{\sqrt{[n]_q!}}\sum_{k=0}^{n}\left[\begin{array}{c}n\\k\end{array}\right]_q z^k,\quad n=0,1,2\cdots,
\eea
one can straightforwardly show   that
\be \label{4qRogersdiff}
D_z^q\psi_n(z;q)= \sqrt{[n]_q}\psi_{n-1}(z;q)
\ee
with the property, for $n=0, 1, 2, \cdots,$
\be
(D_z^q)^{n+1}\psi_n(z;q) = 0\quad\mbox{ and }\quad (D_z^q)^{k}\psi_n(z;q) \neq 0 \quad\mbox{ for any  }
k<n+1.
\ee
Therefore, it follows from  the equations (\ref{4qRogersttr}) and (\ref{4qRogersnorm})
that the polynomials $\{\psi_n(z;q)\; n= 0, 1, 2, \cdots\}$
satisfy the following three-term recurrence relation
\be \label{4qRogersttr1}
 \sqrt{[n+1]_q}\,\psi_{n+1}(z;q)=(1+z)\psi_n(z;q)-z(1-q) \sqrt{[n]_q}\,\psi_{n-1}(z;q).
\ee
By manipulating the  expressions (\ref{4qRogersdiff}) and (\ref{4qRogersttr1}) we obtain the following
  $q-$difference equation
\be \label{4qRogersttr2}
\Big((1+z)-(1-q)z\;D_z^q\Big)\psi_n(z;q)=  \sqrt{[n+1]_q}\,\psi_{n+1}(z;q)
\ee
and the  creation and  annihilation operators  as
\be
A^\dag = 1+z-(1-q)z\;D_z^q \quad\mbox{and}\quad A = D_z^q,
\ee
respectively,
while
 the number operator $N$ formally acts  on the state $\psi_{n}(z;q)$ as
\be \label{4qRogersnb}
N\,\psi_n(z;q)= n\,\psi_{n}(z;q).
\ee

To sum up, the following  relations are in order in this context:
\bea
N\psi_n(z;q)&=& n\,\psi_{n}(z;q),\\
A^\dag\psi_n(z;q)&=&  \sqrt{[n+1]_q}\,\psi_{n+1}(z;q),\\
A\psi_n(z;q)&=&  \sqrt{[n]_q}\,\psi_{n-1}(z;q),\\
A^\dag A\psi_n(z;q)&=&  [n]_q\psi_{n}(z;q) = [N]_q\psi_{n}(z;q),\\
AA^\dag\psi_n(z;q)&=&  [n+1]_q\psi_{n}(z;q) = [N + 1]_q\psi_{n}(z;q)
\eea
and the set of polynomials $\{\psi_n(z;q)\;|n = 0, 1, 2,
\cdots \}$ provides a basis for a realization of the $q$-deformed
harmonic oscillator algebra given by
\be \label{4qHarm}
AA^\dag - q A^\dag A=1,\quad [ N, \;A]=-A ,\quad [N,\;A^\dag]=A^\dag.
\ee

\section{New $q-$Hermite polynomials and oscillator algebra}
Let  $\mathcal{O}(D_R)$ be a set of holomorphic
functions  defined on $D_R=\{z\in {\bf C}:|z|<R\}.$ \\
{\bf Definition 1} \it
The $q-$addition $\op $ is defined as follows
\bea
\label{addition}
 \big(x\op  y\big)^n:&=&(x+y)(x+q y)\ldots (x+q^{n-1}y)\cr
&=&\sum_{k=0}^n{n
\atopwithdelims[] k}_q q^{({}^k_2)}x^{n-k}y^k
,\quad n\geq 1,\quad \big(x\op  y\big)^0:=1,
\eea
while the $q-$subtraction $\om$ is given by the relation
\be
\label{additionm}
 \big(x\om  y\big)^n:= \big(x\op (-y)\big)^n.
\ee
\rm
Consider a function  $F$
\bea
F:D_R  \longrightarrow{\bf  C},\,z \longmapsto \sum_{n=0}^\infty c_n z^n,
\eea
and   define by $F(x\op  y)$  the series
\bea
F(x\op  y):=\sum_{n=0}^\infty c_n(x\op  y)^n.
\eea
We immediately obtain the following rules for the product of two exponential functions
\be
e_q (x) E_q (y) = e_q ( x \op y )
\ee
\be
e_q (x) e_{\qi} (y) = e_q ( x \op y ),
\ee
where
\be
e_q (x) := \sum_{n=0}^{\infty} \frac{1}{[n]_q!}x^n
\ee
\be
e_{\qi} (x) := \sum_{n=0}^{\infty} \frac{1}{[n]_{\qi}!}x^n
\ee
\be
E_q (x) := \sum_{n=0}^{\infty} \frac{q^{n(n-1)/2}}{[n]_q!}x^n.
\ee
Let us introduce the new $q-$Hermite polynomial $H_n^q (x) $ as follows:
\be
\label{functiongeneratrice}
e_q( [2]_q tx \ominus_{q,q^2} t^2 ) = e_q([2]_q tx)E_{q^2} (- t^2 ) := \sum_{n=0}^{\infty} \frac{H_n^q (x) }{[n]_q!}t^n,
\ee
where
\be
(a\ominus _{q, q^{2}}b )^n := \sum_{k=0}^{n}   \frac{[n]_q!}{[n-k]_q! [k]_{q^{2} }!}
(-1)^k q^{k(k-1)} a^{n-k} b^{k},\quad (a\ominus_{q, q^{2}} b )^0  := 1.
\ee
Performing the $q-$derivative \cite{Janga} of   both sides
of  eq.(\ref{functiongeneratrice}) with respect to $x$, one obtains
\bea
\label{lowering}
D_x^q H_n^q(x)=[2]_q[n]_q H_{n-1}^q(x),
\eea
where the $q-$derivative $D_x^q$ is given in eq.(\ref{jackson})
and  the following formula
\bea
D_x^q (a x \op  b)^n=[n]_q(a x \op  b)^{n-1}
\eea
is used.

{\bf Theorem 1} \it For $u=x^2,$ the following holds:
\be
D_x^qf(x^2) =[2]_qx \dd f(u),
\ee
and more generally
\be
(D_x^q)^n f(x^2 ) = \sum_{k=0}^{\langle n/2 \rangle} a_k^n ([2]_qx)^{n-2k} (\dd)^{n-k} f(u),
\ee
where
\be
a_k^n = q^{(n-2k-1)(n-2k)/2} \frac{[n]_q! [2]_q^k }{[n-2k]_q! [2k]_q!! }
\ee
and  $\langle x \rangle$ implies a Gauss symbol.\\
\rm {\bf Proof.} As $a_k^n$ obeys the recurrence relation
\be
a_k^{n+1} = q^{ n-2k } a_k^n +[2]_q[n-2k+2]_q a_{k-1}^n ,
\ee
we have
\be
(D_x^q)^{n+1} f(x^2 ) = \sum_{k=0}^{\langle n+1/2 \rangle} a_k^{n+1} ([2]_q x)^{n+1-2k} (\dd)^{n+1-k} f(u),
\ee
which proves the Theorem 1.
\rm

The eq.(26) can be rewritten as
\be
a_k^n = q^{(n-2k-1)(n-2k)/2} \frac{[n]_q! }{[n-2k]_q! [k]_{q^2}! },
\ee
where we used
\be
[2]_q[k]_{q^2}=[2k]_q.
\ee
Performing the $q-$differentiation of both sides of
 the eq.(\ref{functiongeneratrice}) with respect to $t$, we have
\bea
\label{raising}
H_{n+1}^q(x)=[2]_q x H_{n}^q(x)-[2]_q [n]_q q^{n-1}H_{n-1}^q(x),\quad n\geq 1
\eea
and $H_0^q (x):=1$ by definition. Using the recurrence relation eq.(\ref{raising}), we get
\bea
H_1^q (x) &=& [2]_q x  \cr
H_2^q (x) &=& [2]_q^2 x^2 -[2]_q  \cr
H_3^q (x) &=& [2]_q^3 x^3 -[2]_q^2 [3]_q x   \cr
H_4^q (x) &=& [2]_q^4 x^4 -[2]_q^2
 [3]_q[4]_q x^2 + q^2  [3]_q[2]_q^2.
\eea
Generally we have the following.\\
{\bf Theorem 2} \it The series form of the $q$-Hermite polynomial is given by
\be
\label{sama:qhermite}
H_n^q (x) = \sum_{k=0}^{\langle \frac{n}{2} \rangle}
 \frac{ (-1)^k q^{k(k-1)} [n]_q!  }{[n-2k]_q! [k]_{q^2}! } ([2]_qx)^{n-2k}.
\ee
\vspace{1cm}
\rm
{\bf Proof.} Expanding the generation function
given in eq.(\ref{functiongeneratrice}) in Maclaurin series, we have
\bea
 e_q([2]_q tx)E_{q^2} (- t^2 ) &=& \sum_{k=0}^{\infty} \frac{([2]_q xt)^k }{[k]_q!} \sum_{s=0}^{\infty} \frac{(-1)^s q^{s(s-1)} }{[s]_{q^2} ! } t^{2s} \cr
 &=& \sum_{k=0}^{\infty} \sum_{s=0}^{\infty}  \frac{(-1)^s q^{s(s-1)} ([2]_q x)^k }{[k]_q! [s]_{q^2} ! } t^{k+ 2s}.
\eea
By substituting
\be
k +2s =n ~\Rightarrow~ s \le \langle \frac{n}{2} \rangle,
\ee
then we have
\be
 e_q([2]_q tx)E_{q^2} (- t^2 ) = \sum_{n=0}^{\infty} \lb \sum_{s=0}^{\langle  \frac{n}{2} \rangle}   \frac{(-1)^s q^{s(s-1)} ([2]_q x)^{n-2s} }{[n-2s]_q! [s]_{q^2} ! }\rb  t^{n},
\ee
which achieves the proof.\\
The alternative expression for the $q-$Hermite polynomial is as follows :
\be
H_n^q (x) := E_{q^2} \lb - \frac{1}{[2]_q^2 } (D_x^q )^2 \rb ( [2]_qx )^n
\ee
or, equivalently,
\be
H_n^q (x) = (-1)^n e_{q^{-2} } (x^2 ) (D_x^q)^n e_{q^2} ( -x^2 )
\ee
or, in a developed form,
\be
H_n^q (x) = ([2]_qx -q^{n-2} D_x^q ) ([2]_q x -q^{n-3} D_x^q ) \cdots ([2]_q x - D_x^q ) ([2]_q x -q^{-1}D_x^q )  1.
\ee
{\bf Definition 2}
\it Let ${\cal H}_F$ be the Hilbert space spanned by the basis vectors
$\{\psi_n^q(x),\, n = 1, 2 ,\ldots \}$ such that
\be
\label{ortho}
\psi_n^q(x):=\frac{H_n^q(x)}{\sqrt{[2]_q^n[n]_q!}}
=\frac{1}{\sqrt{[2]_q^n[n]_q!}}\sum_{k=0}^{\langle \frac{n}{2} \rangle}
\frac{(-1)^k q^{k(k-1)}[n]_q!}{[n-2k]_q![k]_{q^2}!}
\big([2]_q x\big)^{n-2k},\; n=0,1,2,...
\ee
\rm
One can straightforwardly show   that
\be
\label{sama:eqaa}
D_x^q\,\psi_n^q(x)= \sqrt{[2]_q[n]_q}\,\psi_{n-1}^q(x),
\ee
which satisfies the property
\be
(D_x^q)^{n+1}\,\psi_n^q(x) = 0\quad\mbox{ and }\quad (D_x^q)^{k}\,\psi_n^q(x)
 \neq 0 \quad\mbox{ for any  }
k<n+1,\,n=0, 1, 2, \ldots.
\ee
Therefore, it follows from  the equation eq.(\ref{raising})
that the polynomials $\{\psi_n^q(x)\;|
\; n= 0, 1, 2, \ldots\}$
satisfy the following three-term recurrence relation
\be
\label{sama:eqa}
 \sqrt{[2]_q [n+1]_q}\,\psi_{n+1}^q(x)= [2]_q  x\,
\psi_n^q(x)-q^{n-1}
\sqrt{[2]_q [n]_q}\,\psi_{n-1}^q(x)
\ee
from which we deduce the following
  $q-$difference equation
\be
\left(  \sqrt{[2]_q}   \,x-q^{n-1} \frac{1}{\sqrt{[2]_q}}D_x^q\right)
\psi_n^q(x)=\sqrt{ [n+1]_q}\,\psi_{n+1}^q(x)
\ee
and, consecutively, the   annihilation (lowering)
$A$  and creation (raising)
$A^\dag$
operators
 as
\be
\label{loweringandraising}
A =  \frac{1}{\sqrt{[2]_q}   }D_x^q \quad\mbox{ and }\quad A^\dag = \sqrt{[2]_q}   \,x- \frac{q^{N-1}}{\sqrt{[2]_q}}D_x^q,
\ee
respectively, as well as
\be
N\psi_n^q(x)= n\psi_{n}^q(x).
\ee
\\
{\bf Theorem 3}
The operators $A,\,A^\dag$ and $N$
obey the following commutation relations
\bea
\label{sama:algebra}
 AA^\dag-qA^\dag A={\bf 1},\quad
[ N, \;A]=-A ,\quad
 [N,\;A^\dag]=A^\dag.
\eea
Furthermore, as matter of the compilation of operator actions on the states $\psi_n^q(x)$, we list the following:
\bea
N\psi_n^q(x)&=& n\psi_n^q(x),\\
A^\dag\psi_n^q(x)&=&  \sqrt{  [n+1]_q}\,\psi_{n+1}^q(x),\\
A\psi_n^q(x)&=&  \sqrt{ [n]_q}\,\psi_{n-1}^q(x),\\
\label{adaga}
A^\dag A\psi_n^q(x)&=&  [n]_q\psi_n^q(x) =  [N]_q\psi_n^q(x),\\
AA^\dag\psi_n^q(x)&=&    [n+1]_q\psi_n^q(x) =   [N + 1]_q\psi_n^q(x).
\eea
Finally,  using
the explicit realization of
the lowering and raising operators
we arrive at
the second order $q-$differential equation
obeyed by the new $q-$Hermite polynomials:
\bea
\label{sama:dhermitedifference}
\Big((D_x^{q})^2-[2]_q \,x \,q^{2-n}D_x^{q}+[2]_q[n]_q\, q^{2-n}\Big)H_{n}^q(x)= 0.
\eea
In the limit when $q\to 1$, eq.(\ref{sama:dhermitedifference})
is reduced to the second order differential equation satisfied
 by the  Hermite polynomials   eq.(\ref{second}).\\
\section{Relevant operator properties}
{\bf Theorem 4} \it
 The operators $Q:=[2]_q^{-1/2}(A^\dag+A)$ and
$P:=i[2]_q^{-1/2}(A^\dag-A),$ defined on the Fock space
 $\mathcal{H}_F,$ are bounded and, consequently,
self-adjoint  if $q<1$. If $q>1$, they are not self-adjoint, but
have a one-parameter family of
self-adjoint extensions. \\
\rm
{\bf Proof.}
The matrix elements of  the operator
$ [2]_q^{-1/2}(A^\dag+A)$ on the Fock space basis vector
$|n\rangle=\psi_n(x)$ are given by
\bea
\langle m|Q|n\rangle&:=&
\langle m| [2]_q^{-1/2}(A^\dag+A)|n\rangle\cr
&=&
b_n\,\delta_{m,n+1}+
b_{n-1}\,\delta_{m,n-1}, \;n,\;m=0,
\;1,\;2,\;\cdots
\eea
while the matrix elements of the operator
$i[2]_q^{-1/2}(A^\dag-A)$  are given by
\bea
\langle m|P|n\rangle&:=&
\langle m|i[2]_q^{-1/2}(A^\dag-A)|n\rangle\cr
&=&
i b_n\,\delta_{m,n+1}-i b_{n-1}\,
\delta_{m,n-1},\; n,\;m=0,\;1,\;2,\;\cdots
\eea
where  $b_{n}=\sqrt{[2]_q^{-1}[n+1]_q}$. Then,
the position operator $Q$ and momentum operator $P$
   can be represented by the two following symmetric Jacobi matrices, respectively,
\bea\label{jacobir}
M_Q= \left(\begin{array}{cccccc}0&
b_{1}&0&0&0&\cdots\\b_{1}
&0&b_{2}&0&0&\cdots\\0
&b_{2}&0&b_{3}&0&\cdots\\\vdots&\ddots&
\ddots&\ddots&\ddots&\ddots
       \end{array}\right)
\eea
and
\bea\label{jacobic}
M_P= \left(\begin{array}{cccccc}0&-ib_{1}
&0&0&0&\cdots\\ib_{1}
&0&-ib_{2}&0&0&\cdots\\0&ib_{2}&0
&-ib_{3}&0&\cdots\\\vdots&
\ddots&\ddots&\ddots&
\ddots&\ddots
       \end{array}\right).
\eea
\begin{itemize}
\item Suppose  $q<1$, then,
\be
 \left|b_n\right|=
\left(\frac{1-q^{n+1}}{1-q^2}\right)^{1/2}<
\left(\frac{1}{1-q^2}\right)^{1/2},\; \forall \;n\geq1.
\ee
Therefore, the Jacobi matrices  in
eq.(\ref{jacobir}) and eq.(\ref{jacobic}) are
bounded and self-adjoint (Theorem 1.2., Chapter
VII in Ref. \cite{Berezanskii}). Thus,
$[2]_q^{-1/2}(A^\dag+A)$ and
$i[2]_q^{-1/2}(A^\dag-A)$
 are bounded and, consequently, self-adjoint.
\item Suppose $q>1,$ then
\be
 \lim_{n\to\infty}b_n=\lim_{n\to\infty}
\left(\frac{1-q^{n+1}}{1-q^2}\right)^{1/2}=\infty.
\ee
Considering the series
$\sum_{n=1}^\infty 1/b_n$, we obtain
\be
\overline{\lim_{n\to\infty}}\left(\frac{1/b_{n+1}}{1/b_n}\right)=
\overline{\lim_{n\to\infty}}\left(\frac{1-q^{n+1}}{1-q^{n+2}}
\right)^{1/2}= q^{-1/2}<1.
\ee
This ratio test  leads to the conclusion  that the series
$\sum_{n=1}^\infty 1/b_n$ converges.
Moreover, $1-2q+q^2=(1-q)^2\geq0
\Longrightarrow q^{-1}+q\geq2$. Hence,
\bea
&& 0\leq\frac{1}{1-q^2}\left(1-
q^{n+1}(q+q^{-1})+q^{2n+2}
\right)^{1/2}
\leq\left(1-2q^{n+1}+q^{2n+2}\right)^{1/2}
\frac{1}{1-q^2}\cr
\Leftrightarrow&&0\leq
\left(\frac{1-q^{n+2}}{1-q^2}\right)^{1/2}
\left(\frac{1-q^{n}}{1-q^2}\right)^{1/2}\leq
 \frac{1-q^{n+1}}{1-q^2}  \cr
\Leftrightarrow&&0\leq
\left(\frac{1-q^{n+2}}{1-q^2}\right)^{1/2}
\left(\frac{1-q^{n}}{1-q^2}\right)^{1/2}
\leq  \frac{1-q^{n+1}}{1-q^2}  \cr
\Leftrightarrow &&
0\leq b_{n-1}b_{n+1}\leq b_n^2.
\eea
Therefore, the Jacobi matrices  in eq.(\ref{jacobir}) and eq.(\ref{jacobic}) are not self-adjoint (Theorem 1.5., Chapter VII in Ref. \cite{Berezanskii})
but have each a one-parameter family of self-adjoint extensions instead. This
means that their deficiency subspaces are one-dimensional.  The proof is thus achieved.
\end{itemize}
The following statement holds.\\
{\bf Proposition 1} \it
\begin{itemize}
\item The vectors $|n\rangle$
are eigen-vectors of the  $q-$deformed Hamiltonian $H^q:=\frac{1}{[2]_q}(AA^\dag+A^\dag A)$ with respect to the eigenvalues
\be
\label{mecaprop1}
 E_n^q= \frac{1}{[2]_q}\Big([n]_q+ [n+1]_q\Big),
\ee
\item The mean values of $Q$ and $P$ in the states $|n\rangle$ are zero while
their variances  are given by
\be
 (\Delta  Q)_{|n\rangle}^2 =(\Delta P)_{|n\rangle}^2=  E_n^q,
\ee
where $(\Delta X)_n^2=\langle X^2\rangle_n-\langle X\rangle_n^2$
with $\langle X\rangle_n=\langle n|X|n\rangle$.
\item The position-momentum uncertainty relation is given by
\be
\label{mecaprop4}
 (\Delta Q)_{|n\rangle}(\Delta P)_{|n\rangle}=  E_n^q,
\ee
which is reduced, for the vacuum state, to the expression
\be
\label{mecaprop5}
 (\Delta Q)_{|0\rangle}(\Delta P)_{|0\rangle}=  \frac{1}{1+q}.
\ee
\end{itemize}
\rm
In the limit when $q\to 1,$
one recovers the uncertainty
relation for the non deformed harmonic oscillator.

\section{Coherent states}
{\bf Definition 3} \it
The coherent states of the Barut-Girardello type for the
algebra eq.(\ref{sama:algebra}) in the Fock space ${\cal H}_F$ are defined as
\be
\label{sama: def}
 |z\rangle := {\cal
N}_q^{-1/2}(|z|^2)\sum_{n=0}^{\infty}
\frac{z^n}{\sqrt{[n]_q!}}|n\rangle,
\quad z\in D_q,
\ee
where ${\cal N}_q$ is the normalization factor given by
\be
\label{netsa}
{\cal N}_q(x):=\sum_{n=0}^{\infty}
\frac{x^n}{[n]_q!},
\ee
and
\bea
D_q:=\{z\in C: |z|<R\}\;\; \text{ with } \;\; R=
\left\{\begin{array}{ll}\frac{1}{1-q} &
\text{ if } \;0 <q <1,\\
\infty &\text{ if }\; q>1.\end{array}\right.
\eea
$R$ is the convergence radius of the series eq.(\ref{netsa}).\\
\rm These states exhibit a series of properties as follows.\\
 {\bf Proposition 2} \it
The coherent states, eq.(\ref{sama: def}), are normalized.
\\
\rm
{\bf Proof.}
The product of two coherent states
 $ |z\rangle$ and $ |z'\rangle$ is given by
\be
\label{CSn}
\langle z' |z\rangle = {\cal
N}_q^{-1/2}(|z|^2){\cal
N}_q^{-1/2}(|z'|^2){\cal
N}_q(z\bar{z'}).
\ee
For $z=z',\;\langle z\vert z\rangle = 1$ showing that
the states (eq.(\ref{sama: def})) are normalized.\\
{\bf Proposition 3} \it The coherent states defined in
eq.(\ref{sama: def}) are continuous  in their label z.\\
\rm
{\bf Proof.}
\be
\left\| \vert z\rangle - \vert
z'\rangle\right\|^2 = 2\left(1 -
Re(\langle z\vert z'\rangle)\right).
\ee
So,
$
\left\|\vert
z\rangle - \vert z'\rangle
\right\|^2 \to  0$ as $|z-z'|\to 0$,
since $\langle z\vert z'\rangle\to 1$ as $\vert z - z'\vert \to   0.
$ This completes the proof.\\
Now before proving that these CS solve the identity, some additional definitions and results deserve investigation.\\
{\bf Lemma 1}\it
\be
\label{sama:lemma1}
  \int_0^{\infty} x^n\Big[\frac{d}{d x}\Big]_q \mathcal{N}_q^{-1}(x) = -
\int_0^{\infty} x^n \mathcal{N}_q^{-1}(q x) d_q x.
\ee
\rm {\bf Proof.} Let us introduce the
deformed derivative $ d_q x$ as follows
\bea\label{Kaderiva}
 \Big[\frac{d}{dx}\Big]_q f(x):= \frac{f(x)-f(q x)}{(1-q)x}.
\eea
Then,
\be
\Big[\frac{d}{d x}\Big]_q x^n =[n]_q x^{n-1},
\ee
and
\be
\Big[\frac{d}{d x}\Big]_q
 \mathcal{N}_q(x)= \mathcal{N}_q(x).
\ee
By applying   the Leibniz rule   on the relation
 $\mathcal{N}_q(x)\mathcal{N}_q^{-1}(x) = 1$, we
obtain
\be
 \Big[\frac{d}{d x}\Big]_q \Bigl(\mathcal{N}_q (x)
\mathcal{N}_q^{-1}(x)\Bigr)=\Big[\frac{d}{d
x}\Big]_q \mathcal{N}_q(x)\cdot
\mathcal{N}_q^{-1}(q x)+
 \mathcal{N}_q(x)\cdot\Big[\frac{d}{d x}\Big]_q
\mathcal{N}_q^{-1}(x) = 0
\ee
furnishing
\be
\label{pas}
\Big[\frac{d}{d x}\Big]_q
\mathcal{N}_q^{-1}(x)= -
\mathcal{N}_q^{-1}(q x).
\ee
By replacing eq.(\ref{pas}) in the left hand
side of eq.(\ref{sama:lemma1}), the proof is achieved.\\
{\bf Definition 4} \cite{Kac} \it The improper $q-$integral of
$f(x)$ on the interval $[0,+\infty)$ is
defined to be
\be
\label{sama:conv}
\int_0^{\infty} f( x) d_q x:=\sum_{j=-\infty}^{+\infty}\int_{q^{j+1}}^{q^{j}} f( x) d_q x
\ee
if $0 < q <1,$ or
\be
\label{sama:convs}
\int_0^{\infty} f( x) d_q x:=\sum_{j=-\infty}^{+\infty}
\int_{q^j}^{q^{j+1}}f( x) d_q x
\ee
if $q>1$.\\
\rm
 In  \cite{Kac}, Kac  and Cheung  proved that the improper $q-$integral,
defined in eq.(\ref{sama:conv}) and eq.(\ref{sama:convs}), converges if
$x^\alpha f(x)$ is bounded in a neighborhood of $x=0$ for some $\alpha <1$ and for
sufficiently larger $x$ with  some $\alpha >1$.\\
{\bf Definition 5} \it For any $n>0$, the $q-$analogue of the gamma function, called the {\it $q-$gamma function}, is
defined as:
\bea
\Gamma_q(n):&=&\int_0^{\infty}x^{n-1}\mathcal{N}_q^{-1}(q x) d_q x\cr
&=&\frac{(q;q)_\infty}{(q^n;q)_\infty}(1-q)^{1-n}.
\eea
{\bf  Lemma 2} \it
The following result holds
\be
\label{tare}
\Gamma_q(n+1)= [n]_q!=\int_0^{\frac{1}{1-q}} x^n
\mathcal{N}_q^{-1}(q x) d_q x.
\ee
\rm
{\bf Proof.} By using the $q-$Jackson's integration, i.e
\be
\int_0^{a} x^n f(x):=(1-q)a\sum_{k=0}^\infty q^kf(aq^k),
\ee
the left hand side of (\ref{tare}) becomes
\bea
\label{tar}
\int_0^{\frac{1}{1-q}} x^n \mathcal{N}_q^{-1}(q x) d_q x&= &\sum_{k=0}^\infty q^k\left(\frac{q^k}{1-q}\right)^n (q^{k+1};q)_\infty\cr
&=&\left(\frac{1}{1-q}\right)^n\sum_{k=0}^\infty q^{(n+1)k} (q^{k+1};q)_\infty
\cr
&=& \frac{(q;q)_\infty}{(1-q)^n} \sum_{k=0}^\infty \frac{q^{(n+1)k}}{ (q ;q)_k}\cr
&=& \frac{(q;q)_\infty}{(1-q)^n}\frac{1}{(q^{1+n};q)_\infty}\cr
&=&[n]_q!.
\eea
Now, applying the
 formula of the integration by parts \cite{Kac}, i.e.
\be
\int_0^{\infty}u(x)\cdot \Big[\frac{d}{d x}\Big]_q v(x) =
\int_0^{\infty}\Big[\frac{d}{d x}\Big]_q (u(x)\cdot v(x))-\int_0^{\infty}
\Big[\frac{d}{d x}\Big]_q u(x)\cdot v(q  x),
\ee
to the function $x^n \mathcal{N}_q^{-1}(q x) $
and  using the   Lemma 1, we get
\bea
\int_0^{\infty} x^n \mathcal{N}_q^{-1}(q x) d_q x
 &= &-
 \int_0^{\infty} x^{n}
\Big[\frac{d}{d x}\Big]_q \mathcal{N}_q^{-1}(x)\cr
&=&[n]_q  \int_0^{\infty}
x^{n-1} \mathcal{N}_q^{-1}(q x) d_q x\cr
&=&[n]_q [n-1]_q\int_0^{\infty}
x^{n-2} \mathcal{N}_q^{-1}(q x) d_q x\cr
&\ldots&\cr
&=& [n]_q!\,,
\eea
where $x^n\mathcal{N}_q^{-1}(x)\big|_0^\infty=0,
\;\mathcal{N}_q (\infty):=
\infty$ and $ \mathcal{N}_q (0):=1.$  \\
All tools are now ready to prove the following property exhibited by the defined CS.\\
{\bf Proposition 4} \it  The coherent states (eq.(\ref{sama: def})) solve the unity,
\bea
\label{sama:solve}
 \int_{D_q}d\mu(\bar{z},z)|z\rangle\langle z|={\bf 1},
\eea
where the measure $d\mu(\bar{z},z)$ is given by
\bea
\label{measu2}
d\mu(\bar{z},z)=\left((1-q)|z|^2;q\right)_\infty\left(
q(1-q)|z|^2;q\right)_\infty^{-1}\frac{d^2z}{2\pi },
\eea
if $0 < q <1,$ and
\bea
\label{measu2}
d\mu(\bar{z},z)&=&\frac{|1-q| }{2\pi }
\frac{\mathcal{N}_q(|z|^2)}{\mathcal{N}_q(q|z|^2)}\sum_{k = 0}^{\infty}  |z|^2
\Bigg\{ \delta \left(|z|^2 - \frac{q^k}{|1-q|}\right)  +
\delta\left(|z|^2-\frac{q^{-k+1}}{|1-q|}
\right)\Bigg\}d^2z,
\eea
if $q>1.$\\
\rm
{\bf Proof.} From the left hand side
of eq.(\ref{sama:solve}), we have
\be
\label{resol}
  \int_{D_q}d\mu(\bar{z},z)|z\rangle\langle z|=
\sum_{n,m=0}^\infty\frac{|n\rangle\langle m|}{\sqrt{
[n]_q![m]_q!}}\int_{D_q}\frac{\bar{z}^m z^n}
{\mathcal{N}_q(|z|^2)}d\mu(\bar{z},z).
\ee
Upon passing to polar coordinates,
$z=\sqrt x\;e^{i\theta}$, $d\mu (\bar z,z)= d\omega_q (x)d\theta$
where $0\leq\theta\leq2\pi$, $0<x< R$
 and $\omega_q $ is a positive valued
function, the integral in the left hand side of
eq.(\ref{resol}) is equivalent to
the   Hausdorff power moment problem when $0<q<1$
\be
\label{sama:KaMoment}
 \int_0^{R}x^n\;\frac{2\pi d\omega_q(x)}
{\mathcal{N}(x)}=   [n]_q!,\quad n= 0,\; 1,\; 2,\;\cdots
\ee
 or to the Stieltjes power moment problem when $ q>1$
\be \label{KaMoment}
 \int_0^{\infty}x^n\;\frac{2\pi
d\omega_q(x)}{\mathcal{N}(x)}=
  [n]_q\,!,\quad n= 0,\; 1,\; 2,\;\cdots
\ee
Defining
\be
{\tilde \omega}_q(x):=
\frac{ \omega_q(x) }{{\cal N}_q(x)},
\ee
$\bullet$ If $0 <q < 1$, the moment
problem (eq.(\ref{sama:KaMoment})) is equivalent to
\be
 2\pi \int_0^{\frac{1}{1-q}}x^n\;
 {\tilde \omega}_q(x) d_q x=
 [n]_q\,!,\quad n= 0,\; 1,\; 2,\;\cdots
\ee
By using the Lemma 2, we arrive at
\be
 {\tilde \omega}_q(x)=\frac{\mathcal{N}_q(x)}{\mathcal{N}_q(q x)}=\left((1-q)|z|^2;q\right)_\infty\left(
q(1-q)|z|^2;q\right)_\infty^{-1}.
\ee
Thus,
\bea
d\mu(\bar{z},z)=\left((1-q)|z|^2;q\right)_\infty\left(
q(1-q)|z|^2;q\right)_\infty^{-1}\frac{d^2z}{2\pi }.
\eea
$\bullet$ For $q>1$, the moment  problem (eq.(\ref{KaMoment})) takes the form
\be
\label{samaMoment}
 2\pi \int_0^{\infty}x^n\;
 {\tilde \omega}_q(x) d_q x=
 [n]_q\,!,\quad n= 0,\; 1,\; 2,\;\cdots
\ee
By using Lemma 1,   Lemma 2 and the
  Jackson integral  \cite{Oney} corresponding to the
deformed derivative (eq.(\ref{Kaderiva})) defined as
\be
\label{integratin}
\int_0^{\frac{\infty}{|1-q|}} f(t)dt_q :=
|1-q|\sum_{k = 0}^{\infty}
\frac{ q^{k}}{|1-q|}
f\left(\frac{q^{k} }{|1-q|}\right)+
\frac{q^{-k+1} }{|1-q|}
 f\left(\frac{q^{-k+1} }{|1-q|}\right),
\ee
the
moment
problem (eq.(\ref{samaMoment})) is reduced to
\bea
    [n]_q\,!&=&2\pi \int_0^{\infty}x^n\;
{\tilde \omega}_q(x) d_q x\cr
&=&\int_0^{\infty}x^n \mathcal{N}_q^{-1}(q x) d_q x \cr
 &= & \sum_{k = 0}^{\infty} q^{k}
\left(\frac{q^{k} }{|1-q|}\right)^n
\mathcal{N}_q^{-1}\left(\frac{q^{k+1} }{|1-q|}\right)+q^{-k+1} \left(\frac{q^{-k+1} }{|1-q|}
\right)^n\mathcal{N}_q^{-1}\left(\frac{q^{-k+2} }{|1-q|}\right)\cr
&=&\sum_{k = 0}^{\infty}
|1-q|y\Bigg\{ \delta
 \left(y - \frac{q^k }{|1-q|}\right) +
\delta\left(y-\frac{q^{-k+1} }{|1-q|}
\right)\Bigg\} y^n\mathcal{N}_q^{-1}(q y).
\eea
So,
\be
{\tilde \omega}_q(x) =\mathcal{N}_q^{-1}
(q x)\sum_{k = 0}^{\infty}
|1-q|x\Bigg\{ \delta \left(x -
 \frac{q^k }{|1-q|}\right) +
\delta\left(x-\frac{q^{-k+1} }{|1-q|}
\right)\Bigg\}
\ee
and the measure
$d \mu(\bar{z},z)$ is given by
\bea
d \mu(\bar{z},z)=
\frac{|1-q|}{2\pi }\frac{\mathcal{N}_q(|z|^2)}{\mathcal{N}_q(q|z|^2)}
 \sum_{k = 0}^{\infty}
 |z|^2\Bigg\{ \delta \left(|z|^2 -
 \frac{ q^k}{|1-q|}\right)  +
\delta\left(|z|^2-\frac{
q^{-k+1}}{|1-q|} \right)\Bigg\}d^2z,
\eea
what achieves the proof.
\section{Concluding remarks}
In this work, we have performed a construction of new $q-$Hermite polynomials and  characterized their main properties.
The corresponding raising and lowering operators are defined. The three-term recursive relation as well as the second-order
differential equation obeyed by these new polynomials have been explicitly derived. Relevant operator actions
as well as  the eigenvalue problem of the deformed oscillator have been investigated. The self-adjointness
of the related position and momentum operators and the uncertainty principle have been studied.
The associated coherent states have been built  and discussed with respect to their main properties.
The induced moment problem has been explicitly solved.

\section*{Acknowledgements}
MNH and SA acknowledge   the Abdus Salam International
Centre for Theoretical Physics (ICTP, Trieste, Italy) for its support through the
Office of External Activities (OEA) - \mbox{Prj-15}. The ICMPA
is also in partnership with
the Daniel Iagolnitzer Foundation (DIF), France.
%
%
\def\JMP #1 #2 #3 {J. Math. Phys {\bf#1},\ #2 (#3)}
\def\JP #1 #2 #3 {J. Phys. A {\bf#1},\ #2 (#3)}


\section*{References}
\begin{thebibliography}{21}
\bibitem{Jimbo}  M. Jimbo, Lett.Math.Phys.{\bf 10 }, 63 ( 1985) ; {\bf 11}, 247 (1986 ) .

\bibitem{Drinfeld} V. Drinfeld, Proc, intern, congress of Mathematicians ( Berkeley, 1986) 78.

\bibitem{AC} M.Arik and D.Coon, \JMP 17 524 1976 .

\bibitem{Macfarlane}  A.J.Macfarlane, \JP 22 4581 1989 .



\bibitem{Biedenharn} L.Biedenharn, \JP 22 L873 1990 .

\bibitem{Jackson1} F. Jackson, Mess.Math. {\bf 38 }, 57 (1909).









\bibitem{Galetti} D. Galleti, A realization of the q-deformed harmonic oscillator: Rogers-Szeg\"o and Stieltjes-Wigert polynomials,
{\it  Braz. J. Phys. } {\bf 33}, 148-157, (2003).
%
\bibitem{Dezire} J.D. Bukweli Kyemba and M.N. Hounkonnou, Characterization of $({\cal R},p,q)$-deformed Rogers-Szeg\"o polynomials: associated quantum algebras, deformed Hermite polynomials and relevant properties, J. Phys. A: Math. Theor. {\bf 45} (2012), 225204.
%
\bibitem{Jagannathan&Sridhar10} R. Jagannathan and R. Sridhar, {$(p,q)$-Rogers-Szeg\"o polynomials and the $(p,q)$-oscillator}, {\it  arXiv: 1005.4309v1 [math.QA]}.
%
\bibitem{Janga} R. Jagannathan and K. Srinivasa Rao, {\rm Two-parameter quantum algebras, twin-basic numbers, and associated generalized hypergeometric series}, {\it arXiv: math/0602613}.
%
\bibitem{Ali&al} S. T. Ali, J.-P. Antoine, J.-P. Gazeau and U. A. Mueller, {\rm Coherent states and their generalizations:
A mathematical overview}, {\it Rev. Math. Phys.} {\bf7} 1013-1104, (1995).
%
\bibitem{Klauder63a} R. J. Klauder, {\rm Continuous-representation theory I. Postulates of
continuous representation theory}, {\it J. Math. Phys.} {\bf 4} 1055-1058, (1963).
%
\bibitem{Klauder63b} R. J. Klauder, {\rm Continuous-representation theory II. Generalized
relation between quantum and classical dynamics}, {\it  J. Math. Phys.} {\bf 4} 1058-1073, (1963).
%
\bibitem{Berezanskii} Ju. M. Berezanski\'{i}, {\it Expansions in Eigenfunctions of Selfadjoint Operators}, (Amer. Math. Soc., Providence, Rhode Island, 1968).
%
\bibitem{Akhiezer} N. I. Akhiezer, {\it The Classical Moment Problem and Some Related Questions in Analysis} (Olivier and Boyd, London, 1965).
%
\bibitem{Tarmakin} J. A. Shohat and J. D. Tamarkin, {\it The Problem of Moments}, (APS, New York, 1943).
%
\bibitem{Oney} S.  Oney, {\rm The Jackson Integral} (2007).
%
\bibitem{Kac} Kac Victor and Cheung Pokman, {\it
Quantum Calculus}, (Springer, New York, 2002).
\end{thebibliography}
\end{document}